\documentclass[a4paper,11pt]{article}
\pdfoutput=1 

\usepackage{jheppub} 

\usepackage[T1]{fontenc} 

\usepackage{mathrsfs}
\usepackage{amsmath}
\usepackage{braket}
\usepackage{dsfont}
\usepackage{hyperref}

\usepackage{slashed} 
\usepackage{enumitem}
\usepackage{physics}

\usepackage{graphicx}
\usepackage{dcolumn}
\usepackage{bm}
\def\p{\textbf{p}}

\def\0{\textbf{0}}
\def\P{\textbf{P}}

\def\x{\textbf{x}}
\def\y{\textbf{y}}

\renewcommand\thefigure{\arabic{figure}}
\newcommand{\gdualn}[1]{\overset{\:{}^{^{{{\neg}}}}}{\smash[t]{#1}}} 

\allowdisplaybreaks

\title{\boldmath The anomalous spin-statistics connection arising from pseudo-Hermiticity}

\author[a]{Yao Bai,}
\emailAdd{202427021017@stu.cqu.edu.cn} 
\affiliation[a]{Department of Physics and Chongqing Key Laboratory for Strongly Coupled Physics,
Chongqing University, Chongqing 401331, China} 
\author[a]{Cheng-Yang Lee,}
\emailAdd{chengyanglee@outlook.com} 
\author[b]{Ruifeng Leng,}
\affiliation[b]{Department of Physics and Center for Field Theory and Particle Physics, \\Fudan University, Shanghai 200438, China}
\emailAdd{lruifeng@fudan.edu.cn} 
\author[a]{Siyi Zhou}
\emailAdd{siyi@cqu.edu.cn}

\abstract{We establish a new spin-statistics theorem for a class of free pseudo-Hermitian quantum field theories whose particles furnish unitary irreducible representations of the Poincar\'{e} group. In this framework, free pseudo-Hermitian fields with integer spin exhibit fermionic statistics, whereas those with half-integer spin exhibit bosonic statistics, opposite to the conventional case. This reversal arises from defining canonical field operators using pseudo-Hermitian conjugation rather than Hermitian conjugation, thereby circumventing the conventional spin-statistics theorem. The free fields retain locality, Lorentz covariance, and unitary evolution. However, interactions may violate unitarity due to the intrinsically non-Hermitian nature of the full Hamiltonian. We discuss potential resolutions to restore unitarity in interacting theories.}

\begin{document}
\maketitle
\flushbottom

\section{Introduction}
\label{sec:intro}

Theorems of physical significance in quantum field theory (QFT) are rare. Most theorems either apply to models in lower-dimensional spacetime or to theories with symmetries that have not been observed in nature. The spin-statistics theorem (SST) is an exception, as it correctly describes the connection between spin and statistics for all observed elementary particles~\cite{Fierz:1939zz,Pauli:1940zz,Feynman:1949hz,PhysRev.110.1450,Pauli1988,Streater:1989vi}. In 3+1 dimensional Minkowski spacetime, the SST can be easily stated: \textit{Integer spin fields are bosonic, while half-integer spin fields are fermionic.}

The SST is derived from the principles of locality and Poincar\'{e} symmetry in QFT. It is therefore a natural consequence of the underlying principles of QFT. While it is commonly assumed that locality and Poincar\'{e} symmetry uniquely imply the standard connection between spin and statistics stated in the preceding paragraph, this assumption is not true. 

The works of LeClair et al.~\cite{LeClair_2007,Robinson:2009xm} and Ahluwalia et al.~\cite{Ahluwalia:2020jkw,Ahluwalia:2022zrm} have shown that it is possible to construct physically well-defined fermionic scalar and bosonic spinor fields. These fields are well-defined in the sense that they are local, respect Poincar\'{e} symmetry and have free Hamiltonians that are Hermitian and positive-definite. The standard SST is circumvented by exploiting the freedom to define non-trivial pseudo-Hermitian field adjoints, while preserving all the features that make the theory physically well-defined.

Here, we extend the above works to establish a theoretical framework for pseudo-Hermitian quantum fields. These theories are pseudo-Hermitian in the sense that the full Hamiltonian $H$ satisfies the pseudo-Hermiticity definition~\cite{Mostafazadeh:2001jk,Mostafazadeh:2001nr,Mostafazadeh:2002id,Mostafazadeh:2008pw}
\begin{equation}
H^{\#}=\eta^{-1}H^{\dag}\eta = H \, , 
\quad
\eta^{\dag}=\eta \,,
\end{equation}
where $\eta$ is a linear and Hermitian intertwining operator. In section~\ref{sec:sst}, we will give a more refined definition of pseudo-Hermiticity in terms of the free quantum fields without referring to the full Hamiltonian.
The pseudo-Hermitian quantum fields furnish the unitary irreducible representation of the Poincar\'{e} group, so they are local and Lorentz covariant. However, they satisfy a new spin-statistics theorem (NSST): \textit{Integer spin fields are fermionic, while half-integer spin fields are bosonic.} The standard SST is circumvented by exploiting the freedom to construct pseudo-Hermitian field adjoints. 



The free Lagrangian densities for pseudo-Hermitian QFTs are non-Hermitian. Nevertheless, the corresponding free Hamiltonian $H_{0}$ and three-momentum $\P_{0}$ are Hermitian. Consequently, the generators of the Poincar\'{e} group remain Hermitian and satisfy the Poincar\'{e} algebra, ensuring the unitarity of the free pseudo-Hermitian theory. In contrast, the full Hamiltonian $H=H_{0}+V$, where $V$ denotes the interaction, and the Belinfante-Rosenfeld tensor are generally non-Hermitian. Our ongoing work aims to construct a fully consistent framework for pseudo-Hermitian interacting field theories.

In the pseudo-Hermitian QFT literature (including $PT$-symmetric theories)~\cite{Bender:2004sa,Bender:2004vn,Bender:2004ej,Novikov:2019ntb,Li:2023kpi,Sablevice:2023odu,Mason:2023sgr,Li:2024xms,Chernodub:2025wga}, 
the formalism proposed by Sablevice and Millington (SM)~\cite{Sablevice:2023odu} is closely related to our construct. Quantum fields in this formalism furnish the pseudo-Hermitian representation of the Poincar\'{e} group, thereby ensuring covariance and locality. Their construct is similar to ours in the sense that they also define a pseudo-Hermitian field adjoint, and the Lagrangian density is pseudo-Hermitian. However, it is different from our construct in one crucial aspect. In the formalism of SM, they did not discuss the connections between spin, statistics, and pseudo-Hermiticity. That is, their pseudo-Hermitian quantum fields are implicitly assumed to obey the standard SST, whereas in our construct, the pseudo-Hermitian field obeys the NSST.


The paper is organized as follows. In section~\ref{sec:sst}, we review the basic construction of the free Hermitian QFT and the SST. Subsequently, we construct the free pseudo-Hermitian QFT and derive the NSST. In section~\ref{sec:pH_QFT}, we construct the pseudo-Hermitian scalar, spinor, and vector fields. In section~\ref{sec:Lagrangian}, we present the Lagrangian formalism for pseudo-Hermitian QFT. We show that the generators of the Poincar\'{e} group obtained from the free Lagrangian density are Hermitian and satisfy the Poincar\'{e} algebra. In section~\ref{sec:conclusion}, we discuss future research directions for pseudo-Hermitian QFT.


\section{Spin-statistics theorem}\label{sec:sst}

We start by reviewing the basic constructs of the Hermitian QFT and the standard SST. Subsequently, we will use results from the Hermitian QFT to construct the pseudo-Hermitian QFT obeying the NSST.

Let $\lambda(x)$ be a free field and $\bar{\lambda}(x)$ be its adjoint associated with a Hermitian QFT. We expand the fields as
\begin{align}
\lambda(x)
&=\int\frac{d^{3}p}{(2\pi)^{3}}\frac{1}{\sqrt{2E_{\p}}}\sum_{\sigma}\left[e^{-ip\cdot x}f(\p,\sigma)a(\p,\sigma)+e^{ip\cdot x}g(\p,\sigma)a^{c\dag}(\p,\sigma)\right] \, ,
\label{eq:field_gen_1} 
\\
\bar{\lambda}(x)
&=\int\frac{d^{3}p}{(2\pi)^{3}}\frac{1}{\sqrt{2E_{\p}}}\sum_{\sigma}\left[e^{ip\cdot x}\bar{f}(\p,\sigma)a^{\dag}(\p,\sigma)+e^{-ip\cdot x}\bar{g}(\p,\sigma)a^{c}(\p,\sigma)\right] \, ,
\label{eq:field_gen_dual_1}
\end{align}
where $f,g,\bar{f},\bar{g}$ are the coefficient functions; $a,a^{c\dag}$ are the annihilation and creation operators for particles and anti-particles, respectively. What makes the theory Hermitian will be specified in due course. For now, let us briefly review the particle contents associated with $a,a^{c\dag}$ and their connections with statistics. Given the free vacuum state $|0\rangle$, the single particle and anti-particle states, with appropriate normalizations, are defined as
\begin{align}
        |\p,\sigma,a\rangle&\equiv\sqrt{2E_{\mathbf{p}}}a^{\dag}(\p,\sigma)|0\rangle\,,\\
        |\p,\sigma,a^{c}\rangle&\equiv\sqrt{2E_{\mathbf{p}}}a^{c\dag}(\p,\sigma)|0\rangle\,.
\end{align}
These states furnish the unitary representation of the Poincar\'{e} group. For massive and massless particles, $\sigma=-j,\cdots,j$ and $\sigma=\pm j$, respectively, where $j=0,\frac{1}{2},1\cdots$ is the spin of the particle. In the 3+1-dimensional Minkowski spacetime, the creation and annihilation operators must satisfy the canonical commutation or anti-commutation relations
\begin{equation}
\left[a(\p',\sigma'),a^{\dag}(\p,\sigma)\right]_{\pm}
=\left[a^{c}(\p',\sigma'),a^{c\dag}(\p,\sigma)\right]_{\pm}=(2\pi)^{3}\delta^{3}(\p'-\p)\delta_{\sigma'\sigma} \,,\label{eq:anti-comm}
\end{equation}
where the signs $-$ and $+$ indicate a commutator and an anticommutator respectively
\begin{equation}
[a,b]_{\pm}\equiv ab\pm ba\,.    
\end{equation}
The particles and the associated fields satisfy the bosonic/fermionic statistics if the creation and annihilation operators satisfy commutation/anti-commutation relations. Whether the operators commute or anti-commute is determined by the SST. 

In the standard Hermitian QFT, the connection between spin and statistics is
determined via the demand of Poincar\'{e} covariance, Hermiticity, and locality. Let us examine these demands. Poincar\'{e} covariance means that $\lambda(x)$ and $\bar{\lambda}(x)$ must transform as
\begin{align}
\text{U}(\Lambda,\alpha)\lambda(x)\text{U}^{-1}(\Lambda,\alpha)
&=\mathcal{D}^{-1}(\Lambda)\lambda(\Lambda x+\alpha) \, , 
\\
\text{U}(\Lambda,\alpha)\bar{\lambda}(x)\text{U}^{-1}(\Lambda,\alpha)
&=\bar{\lambda}(\Lambda x+\alpha)\mathcal{D}(\Lambda) \, ,
\end{align}
where $\text{U}(\Lambda,\alpha)$ is the unitary representation of the Poincar\'{e} transformation in the Hilbert space and $\mathcal{D}(\Lambda)$ is the finite-dimensional representation of the Lorentz group. Therefore, $\bar{\lambda}(x)\lambda(x)$ transforms as a scalar operator. The demand of Hermiticity requires that the scalar operator must be Hermitian
\begin{equation}
\left[\bar{\lambda}(x)\lambda(x)\right]^{\dag} = \bar{\lambda}(x)\lambda(x) \, ,
\label{eq:Herm_bi_1}
\end{equation}
so that the interacting densities constructed from $\lambda(x)$ and $\bar{\lambda}(x)$ are Hermitian. This demand also constrains the relation between the field and its adjoint. For example, if $\lambda(x)$ is a complex scalar field, its adjoint is given by $\bar{\lambda}(x)=\lambda^{\dag}(x)$. If $\lambda(x)$ is the Dirac field, then $\bar{\lambda}(x)=\lambda^{\dag}(x)\gamma^{0}$. 


The SST for Hermitian QFT is encoded in the demand for locality. Here, it suffices for us to consider the equal-time commutator/anti-commutator between $\lambda(t,\x)$ and $\bar{\lambda}(t,\y)$. Using eq.~\eqref{eq:anti-comm}, we obtain
\begin{align}
\left[\lambda(t,\x),\bar{\lambda}(t,\y)\right]_{\pm}&=\int\frac{d^{3}p}{(2\pi)^{3}}\,e^{i\mathbf{p\cdot(x-y)}}\frac{1}{2E_{\p}}M_{\pm}(\p)
\, ,
\label{eq:comm_1}
\end{align}
where
\begin{equation}
M_{\pm}(\p)
\equiv \sum_{\sigma}
\left[f(\p,\sigma)\bar{f}(\p,\sigma)\pm g(-\p,\sigma)\bar{g}(-\p,\sigma)\right]
\, .
\label{eq:M_sum_1}
\end{equation}
The SST states the following. Integer spin fields are bosonic. The equal-time commutator between $\lambda(t,\x)$ and $\bar{\lambda}(t,\y)$ is given by\footnote{Strictly speaking, the equal-time commutator/anti-commutator can also be proportional to $\delta^{(3)}(\x-\y)$ or the derivative of $\delta^{(3)}(\x-\y)$. If $\lambda(x)$ is the Dirac field, then $\left[\lambda(t,\x),\bar{\lambda}(t,\y)\right]_{+}=\gamma^{0}\delta(\x-\y)$.}
\begin{align}
\left[\lambda(t,\x),\bar{\lambda}(t,\y)\right]_{-}=0
    \,,\quad M_{-}(\p)=0\,,\quad j=0,1,\cdots\,. \label{eq:M_m}
\end{align}
Fields of half-integer spin are fermionic, so $\lambda(t,\x)$ and $\bar{\lambda}(t,\y)$ anti-commutes at equal-time
\begin{align}
\left[\lambda(t,\x),\bar{\lambda}(t,\y)\right]_{+}=0
    \,,\quad M_{+}(\p)=0\,,\quad j=\frac{1}{2},\frac{3}{2},\cdots\,. \label{eq:M_p}
\end{align}

In 3+1 dimensional Minkowski spacetime, the SST for Hermitian QFT is a natural consequence of the demand of Poincar\'{e} covariance, Hermiticity and locality. Since these demands are part of the fundamental principles of QFT, the SST is usually assumed to be unique. However, the works of LeClair et al.~\cite{LeClair_2007,Robinson:2009xm} and Ahluwalia et al.~\cite{Ahluwalia:2020jkw,Ahluwalia:2022zrm} have shown otherwise. For complex scalar and spinor fields, these authors have shown that the adjoint field, which leads to eq.~\eqref{eq:Herm_bi_1} is not unique. In fact, one can define a pseudo-Hermitian field adjoint to circumvent the SST and construct fermionic scalar and bosonic scalar fields, while preserving Poincar\'{e} covariance and locality. Here, we extend these works to construct a general class of pseudo-Hermitian QFT and derive the NSST. For this purpose, we have to make use of the SST for Hermitian QFT.

Let $\chi(x)$ and $\gdualn{\chi}(x)$ be the free field and its adjoint that are Poincar\'{e} covariant. Specifically, we take their Poincar\'{e} transformations to be identical to $\lambda(x)$ and $\bar{\lambda}(x)$:
\begin{align}
\text{U}(\Lambda,\alpha)\chi(x)\text{U}^{-1}(\Lambda,\alpha)
&=\mathcal{D}^{-1}(\Lambda)\chi(\Lambda x+\alpha)
\, ,  \\
\text{U}(\Lambda,\alpha)\gdualn{\chi}(x)\text{U}^{-1}(\Lambda,\alpha)
&=\gdualn{\chi}(\Lambda x+\alpha)\mathcal{D}(\Lambda)\label{eq:dchi_trans1}
\, .
\end{align}
Therefore, we can take the expansion of $\chi(x)$ to be
\begin{equation}
\chi(x)=\int\frac{d^{3}p}{(2\pi)^{3}}\frac{1}{\sqrt{2E_{\p}}}\sum_{\sigma}\left[e^{-ip\cdot x}f(\p,\sigma)b(\p,\sigma)+e^{ip\cdot x}g(\p,\sigma)b^{c\dag}(\p,\sigma)\right]
\, , 
\label{eq:field_gen_ns_1}
\end{equation}
with the coefficient functions $f,g$ being identical to $\lambda(x)$~\eqref{eq:field_gen_1}. The expansion for $\gdualn{\chi}(x)$, which is the crucial element needed to derive the NSST, is to be derived. To emphasize that $\chi(x),\gdualn{\chi}(x)$ are distinct from $\lambda(x),\bar{\lambda}(x)$, we have introduced a new set of creation and annihilation operators $b$ and $b^{c\dag}$ to be associated with them. These operators must also satisfy the canonical commutation/anti-commutation relations
\begin{align}
    \left[b(\p',\sigma'),b^{\dag}(\p,\sigma)\right]_{\pm}=\left[b^{c}(\p',\sigma'),b^{c\dag}(\p,\sigma)\right]_{\pm}=(2\pi)^{3}\delta_{\sigma'\sigma}\delta^{(3)}(\p'-\p) \, .
\end{align}

In the pseudo-Hermitian construct, we do not take the adjoint of $\chi(x)$ to be $\bar{\chi}(x)$. Doing so would make the fields Hermitian and obey the standard SST. Instead, we define $\gdualn{\chi}(x)$ using the pseudo-Hermitian conjugation\footnote{Our construct is closely related to the field adjoint defined by SM~\cite{Sablevice:2023odu}. Given a field $\hat{\psi}(x)$, SM  defined its adjoint to be~$\hat{\tilde{\psi}}(x)\equiv\hat{\eta}^{-1}\hat{\psi}(x^{\eta})\hat{\eta}\pi$, where $\pi$ is a finite-dimensional Hermitian matrix and $x^{\eta}$ is the transformation on $x$ induced by $\hat{\eta}$. For example, if $\hat{\eta}$ is parity, then $x^{\eta}=(t,-\x)$ where it is understood that $x=(t,\x)$. If $\hat{\psi}=(\hat{\psi}_{1},\hat{\psi}_{2},\cdots,\hat{\psi}_{n})$ is a field multiplet consists of $\hat{\psi}_{i}$ ($1\leq i\leq n$) fields, then the dimension of $\pi$ is $n\times n$ (See~\cite{Sablevice:2023odu} for more details). 
}
\begin{equation}
\gdualn{\chi}(x)\equiv\eta^{-1}\bar{\chi}(x)\eta
\, , \quad 
\eta^{\dag} = \eta \, , \label{eq:ph_adj}
\end{equation}
where $\eta$ is a linear and Hermitian operator and
\begin{equation}
\bar{\chi}(x)=\int\frac{d^{3}p}{(2\pi)^{3}}\frac{1}{\sqrt{2E_{\p}}}\sum_{\sigma}\left[e^{ip\cdot x}\bar{f}(\p,\sigma)b^{\dag}(\p,\sigma)+e^{-ip\cdot x}\bar{g}(\p,\sigma)b^{c}(\p,\sigma)\right] \,.
\end{equation}
The pseudo-Hermitian field adjoint~\eqref{eq:ph_adj} is a generalization of the Hermitian field adjoint. The solution for $\eta$ is constrained by Lorentz covariance and locality. 
obeying the NSST.

The field $\bar{\chi}(x)$ is Poincar\'{e}-covariant for the same reason that $\bar{\lambda}(x)$ is Poincar\'{e}-covariant. Therefore, we may express the Poincar\'{e} transformation for $\gdualn{\chi}(x)$ as
\begin{align}
    \text{U}(\Lambda,\alpha)\gdualn{\chi}(x)\text{U}^{-1}(\Lambda,\alpha)&=\eta^{-1}_{\Lambda,\alpha}\bar{\lambda}(\Lambda x+\alpha)\eta_{\Lambda,\alpha}\mathcal{D}(\Lambda) \,, \label{eq:dchi_trans2}
\end{align}
where $\eta_{\Lambda,\alpha}\equiv U(\Lambda,\alpha)\eta U^{-1}(\Lambda,\alpha)$. Comparing eq.~\eqref{eq:dchi_trans1} with eqs.~\eqref{eq:ph_adj} and~\eqref{eq:dchi_trans2}, we see that for $\gdualn{\chi}(x)$ to be Poincar\'{e}-covariant, $\eta$ must commute with $\text{U}(\Lambda)$. Since $\eta$ is linear, it only acts on $b^{\dag},b^{c}$. To determine $\eta$, we need to know how it acts on $b^{\dag},b^{c}$. Since $b^{\dag},b^{c}$ furnish unitary irreducible representation of the Poincar\'{e} group, the Poincar\'{e} transformations for $b^{\dag}$ and $b^{c}$ are
\begin{align}
\text{U}(\Lambda,\alpha)b^{\dag}(\p,\sigma)\text{U}^{-1}(\Lambda,\alpha)
&=e^{-ip\cdot\alpha}\sum_{\sigma'}D_{\sigma'\sigma}(W(\Lambda,p))b^{\dag}(\mathbf{\Lambda p},\sigma')
\, , \\
\text{U}(\Lambda,\alpha)b^{c}(\p,\sigma)\text{U}^{-1}(\Lambda,\alpha)
&=e^{ip\cdot\alpha}\sum_{\sigma'}D^{*}_{\sigma'\sigma}(W(\Lambda,p))b^{c}(\mathbf{\Lambda p},\sigma')
\, , 
\end{align}
where $D(W)$ is the unitary irreducible representation of the little group $W$~\cite[Sec.~2]{Weinberg:1995mt}. The simplest non-trivial solution for $\eta$ is that it is an internal symmetry operator that leaves $\p$ and $\sigma$ of $b^{\dag},b^{c}$ invariant.
Therefore,
\begin{align}
\eta^{-1}b^{\dag}(\p,\sigma)\eta
&=e^{-i\theta}b^{\dag}(\p,\sigma) \, , \\
\eta^{-1}b^{c}(\p,\sigma)\eta
&=e^{i\theta^{c}}b^{c}(\p,\sigma) \, , \label{eq:eta_cd}
\end{align}
where $\theta,\theta^{c}$ are constant real phases to be determined. Only the relative phase between $b^{\dag}(\p,\sigma)$ and $b^{c}(\p,\sigma)$ is meaningful, so without loss of generality, we can set $\theta=0$ and take $\gdualn{\chi}(x)$ to be
\begin{equation}
\gdualn{\chi}(x)=\int\frac{d^{3}p}{(2\pi)^{3}}\frac{1}{\sqrt{2E_{\p}}}\sum_{\sigma}\left[e^{ip\cdot x}\bar{f}(\p,\sigma)b^{\dag}(\p,\sigma)+e^{i\theta^{c}}e^{-ip\cdot x}\bar{g}(\p,\sigma)b^{c}(\p,\sigma)\right] \, .
\end{equation}
At equal-time, the commutator/anti-commutator between $\chi(t,\x)$ and $\gdualn{\chi}(t,\y)$ is given by
\begin{equation}
\left[\chi(t,\x),\gdualn{\chi}(t,\y)\right]_{\pm}=\int\frac{d^{3}p}{(2\pi)^{3}}\frac{1}{2E_{\p}}e^{i\mathbf{p\cdot(x-y)}}N_{\pm}(\p,\theta^{c}) \, , 
\end{equation}
where
\begin{equation}
N_{\pm}(\p;\theta^{c}) \equiv \sum_{\sigma} 
\left[f(\p,\sigma)\bar{f}(\p,\sigma)\pm e^{i\theta^{c}}g(-\p,\sigma)\bar{g}(-\p,\sigma)\right]
\, .
\end{equation}
From eqs.~\eqref{eq:M_m}-\eqref{eq:M_p}, we see that the spin-sum $N_{\pm}(\p,\theta^{c})$ admits two solutions for which $\chi,\gdualn{\chi}$ is local:
\begin{alignat}{2}
    N_{\pm}(\p;n\pi)&=M_{\pm}(\p)=0\,,\quad e^{i\theta^{c}}=e^{in\pi}=1\,,&\quad n&\in2\mathbb{Z}\,,\label{eq:N1}\\
    N_{\pm}(\p;n'\pi)&=M_{\mp}(\p)=0\,,\quad e^{i\theta^{c}}=e^{in'\pi}=-1\,,&\quad n'&\in2\mathbb{Z}+1\,.\label{eq:N2}
\end{alignat}
These two solutions, especially the latter, are the main results of this work. Let us examine both solutions separately.
\\

\noindent{\textbf{Trivial solution~\eqref{eq:N1}}  This solution is trivial since $\eta=I$, implying that $\gdualn{\chi}(x)=\bar{\chi}(x)$. Therefore, the theory reduces to the standard Hermitian QFT. The commutator/anti-commutator for $\chi(t,\x)$ and $\gdualn{\chi}(t,\y)$ is equal to the the commutator/anti-commutator for $\lambda(t,\x)$ and $\bar{\lambda}(t,\y)$
\begin{align}
    \left[\chi(t,\x),\gdualn{\chi}(t,\y)\right]_{-}=\left[\lambda(t,\x),\bar{\lambda}(t,\y)\right]_{-}&=0\,,\quad j=0,1,\cdots\,,\\
    \left[\chi(t,\x),\gdualn{\chi}(t,\y)\right]_{+}=\left[\lambda(t,\x),\bar{\lambda}(t,\y)\right]_{+}&=0\,,\quad j=\frac{1}{2},\frac{3}{2},\cdots\,.
\end{align}
\\
\noindent \textbf{Non-trivial solution~\eqref{eq:N2}} This is the main result. When $e^{i\theta^{c}}=-1$, the solution for $\eta$ is given by~\footnote{We can derive the solution for $\eta$ using the fact that it commutes with $b$ and anti-commutes with $b^{c}$.}~\cite{Kapit:2008dp}
\begin{equation}
    \eta=\exp\left[i\pi\int\frac{d^{3}p}{(2\pi)^{3}}\sum_{\sigma}b^{c\dag}(\p,\sigma)b(\p,\sigma)\right] \, .
    \label{eq:eta}
\end{equation}
Therefore, the \textit{commutator/anti-commutator for} $\chi(t,\x)$ \textit{and} $\gdualn{\chi}(t,\y)$ \textit{is equal to the} \textit{anti-commutator/commutator} for $\lambda(t,\x)$ \textit{and} $\bar{\lambda}(t,\y)$:
\begin{align}
    \left[\chi(t,\x),\gdualn{\chi}(t,\y)\right]_{+}&=\left[\lambda(t,\x),\bar{\lambda}(t,\y)\right]_{-}=0\,,\quad j=0,1,\cdots\,,\label{eq:f}\\
    \left[\chi(t,\x),\gdualn{\chi}(t,\y)\right]_{-}&=\left[\lambda(t,\x),\bar{\lambda}(t,\y)\right]_{+}=0\,,\quad j=\frac{1}{2},\frac{3}{2}\cdots\,.\label{eq:b}
\end{align}
Equations~\eqref{eq:f}-\eqref{eq:b} constitute the NSST: \textit{Integer spin fields are fermionic and half integer spin fields are bosonic}. The fields $\chi(x)$ and $\gdualn{\chi}(x)$ associated with eqs.~\eqref{eq:eta}-\eqref{eq:b} have the expansions
\begin{align}
    \chi(x)&=\int\frac{d^{3}p}{(2\pi)^{3}}\frac{1}{\sqrt{2E_{\mathbf{p}}}}\sum_{\sigma}\left[e^{-ip\cdot x}f(\p,\sigma)b(\p,\sigma)+e^{ip\cdot x}g(\p,\sigma)b^{c\dag}(\p,\sigma)\right]\,,\label{eq:chi}\,\\
    \gdualn{\chi}(x)&=\int\frac{d^{3}p}{(2\pi)^{3}}\frac{1}{\sqrt{2E_{\mathbf{p}}}}\sum_{\sigma}\left[e^{ip\cdot x}\bar{f}(\p,\sigma)b^{\dag}(\p,\sigma)-e^{-ip\cdot x}\bar{g}(\p,\sigma)b^{c}(\p,\sigma)\right]\,.\label{eq:chi_adj}
\end{align}
The minus sign in eq.~\eqref{eq:chi_adj} is the crucial feature that makes $\chi(x)$ and $\gdualn{\chi}(x)$ obey the NSST. In this sense, they constitute a non-trivial pseudo-Hermitian QFT where the scalar operator $\gdualn{\chi}(x)\chi(x)$ is non-Hermitian but it is pseudo-Hermitian with respect to $\eta$
\begin{equation}
    \eta^{-1}\left[\gdualn{\chi}(x)\chi(x)\right]^{\dag}\eta=\gdualn{\chi}(x)\chi(x)\,.\label{eq:ph_scalar}
\end{equation}
Despite the non-Hermiticity~\eqref{eq:ph_scalar}, the free theory is actually Hermitian. If the free theory is non-Hermitian, it would not be physically well-defined. As we will show in section~\ref{sec:Lagrangian}, the free generators of the Poincar\'{e} group are Hermitian and satisfy the Poincar\'{e} algebra. However, interacting densities constructed from $\chi(x)$ and $\gdualn{\chi}(x)$ would, in general, be non-Hermitian, thus making the theory truly pseudo-Hermitian.




\section{Pseudo-Hermitian quantum fields}\label{sec:pH_QFT}

In this section, we construct the pseudo-Hermitian scalar, spinor, and vector fields using the formalism presented in the previous section. We will derive the equations of motion, write down their Lagrangian densities, and derive the discrete symmetries, namely charge-conjugation $C$, parity $P$, and time-reversal $T$.

Using the Lagrangian formalism, we show that the pseudo-Hermitian fields satisfy the canonical equal-time commutators/anti-commutators. Many of these results (for scalar and spinor fields) are already known in the literature. We present them here for pedagogy. Moreover, their discrete symmetries have not been analyzed in previous works. We defer the discussions of the free Hamiltonian $H_{0}$ and three momentum $\P_{0}$ to the next section, where we will provide a generic proof that they are Hermitian and positive-definite.

\subsection{Scalar field}

The pseudo-Hermitian scalar field is an anti-commuting complex scalar field constructed by LeClair et al.~\cite{LeClair_2007,Robinson:2009xm}. This is the symplectic fermion theory. The fields $\phi(x)$ and $\gdualn{\phi}(x)$ have expansions
\begin{align}
\phi(x)
&=\int\frac{d^{3}p}{(2\pi)^{3}}\frac{1}{\sqrt{2E_{\p}}}\left[e^{-ip\cdot x}b(\p)+e^{ip\cdot x}b^{c\dag}(\p)\right]
\, , \\
\gdualn{\phi}(x)
&=\int\frac{d^{3}p}{(2\pi)^{3}}\frac{1}{\sqrt{2E_{\p}}}\left[e^{ip\cdot x}b^{\dag}(\p)-e^{-ip\cdot x}b^{c}(\p)\right]\label{eq:dual_phi}
\,.
\end{align}
The minus sign in eq.~\eqref{eq:dual_phi} does not affect their kinematics, so $\phi(x)$ and $\gdualn{\phi}(x)$ satisfy the Klein-Gordon equation
\begin{equation}(\partial^{\mu}\partial_{\mu}+m^{2})\phi=(\partial^{\mu}\partial_{\mu}+m^{2})\gdualn{\phi}=0\,,
\end{equation}
and have the Lagrangian density
\begin{equation}
    \mathcal{L}_{\phi}=\partial^{\mu}\gdualn{\phi}\partial_{\mu}\phi-m^{2}\gdualn{\phi}\phi\,.\label{eq:L_scalar}
\end{equation}

The fields have spin $j=0$, so they are fermionic. From eq.~\eqref{eq:L_scalar}, the conjugate momenta to $\phi(x)$ and $\gdualn{\phi}(x)$ are~\footnote{Note that $\gdualn{\pi} (x)$ is not equal to $\pi (x)$ after applying the $\eta$-conjugation. The minus sign of $\gdualn{\pi} (x)$ comes from anti-commuting fields $\gdualn{\phi} (x)$ and $\phi (x)$.}
\begin{equation}
    \pi=\frac{\partial\mathcal{L}_{\phi}}{\partial\dot{\phi}}=\dot{\gdualn{\phi}}\,,\quad
    \gdualn{\pi}=\frac{\partial\mathcal{L}_{\phi}}{\partial\dot{\gdualn{\phi}}}=-\dot{\phi}\,.
\end{equation}
Direct computations show that the symplectic fermions satisfy the canonical equal-time anti-commutators
\begin{align}
    \left[\phi(t,\x),\phi(t,\y)\right]_{+}&=\left[\pi(t,\x),\pi(t,\y)\right]_{+}=0\label{eq:phi_anti_c1}\,,\\
    \left[\phi(t,\x),\pi(t,\y)\right]_{+}&=i\delta^{(3)}(\x-\y)\,,\label{eq:phi_anti_c2}
\end{align}
and
\begin{align}
    \left[\gdualn{\phi}(t,\x),\gdualn{\phi}(t,\y)\right]_{+}&=\left[\gdualn{\pi}(t,\x),\gdualn{\pi}(t,\y)\right]_{+}=0\,,\\
    \left[\gdualn{\phi}(t,\x),\gdualn{\pi}(t,\y)\right]_{+}&=i\delta^{(3)}(\x-\y)\,.
\end{align}
The remaining anti-commutators identically vanish
\begin{align}
    \left[\phi(t,\x),\gdualn{\phi}(t,\y)\right]_{+}&=\left[\pi(t,\x),\gdualn{\pi}(t,\y)\right]_{+}=0\,,\\
    \left[\phi(t,\x),\gdualn{\pi}(t,\y)\right]_{+}&=\left[\gdualn{\phi}(t,\x),\pi(t,\y)\right]_{+}=0\,.
\end{align}

The discrete symmetries for symplectic fermions can be analyzed in the same way as the scalar bosons. The $C$, $P$, and $T$ transformations for $b$ and $b^{c\dag}$ are
\begin{alignat}{2}
Cb(\p)C^{-1}&=\eta^{*}_{C}b^{c}(\p)\,,\\
Pb(\p)P^{-1}&=\eta^{*}_{P}b(-\p)\,, \\
Tb(\p)T^{-1}&=\eta^{*}_{T}b(-\p)\,,
\end{alignat}
and
\begin{align}
    Cb^{c\dag}(\p)C^{-1}&=\eta^{c}_{C}b^{\dag}(\p)\,, \\
    Pb^{c\dag}(\p)P^{-1}&=\eta^{c}_{P}b^{c\dag}(-\p)\,,\\
    Tb^{c\dag}(\p)T^{-1}&=\eta^{c}_{T}b^{c\dag}(-\p)\,.
\end{align}
The symplectic fermions have the same parity and time-reversal transformations as the scalar bosons. By choosing the intrinsic parity and time-reversal phases to be even, 
\begin{align}
    \eta^{*}_{P}&=\eta^{c}_{P}\,,\\
    \eta^{*}_{T}&=\eta^{c}_{T}\,,
\end{align}
we obtain
\begin{align}
P\phi(x)P^{-1}&=\eta^{*}_{C}\phi(\mathscr{P}x)\,,\\
T\phi(x)T^{-1}&=\eta^{*}_{T}\phi(\mathscr{T}x)\,,
\end{align}
where
\begin{equation}
    \mathscr{P}x=(t,-\x)\,,\quad
    \mathscr{T}x=(-t,\x)\,.
\end{equation}
However, the charge-conjugation symmetry is different. This is because the field adjoint for $\phi$ is $\gdualn{\phi}$ and not $\phi^{*}$. Charge-conjugation is a symmetry when $C$ maps $\phi$ to $\gdualn{\phi}$. By choosing the charge-conjugation phase to be odd
\begin{align}
    \eta^{*}_{C}=-\eta^{c}_{C}\,,
\end{align}
we obtain
\begin{align}
C\phi(x)C^{-1}=-\eta^{*}_{C}\gdualn{\phi}(x)\,,
\end{align}
and 
\begin{equation}
C\mathcal{L}_{\phi}(x)C^{-1}=\mathcal{L}_{\phi}(x) \,.    
\end{equation}


\subsection{Massive spinor field}


The massive pseudo-Hermitian spinor field in the $\left(\frac{1}{2},0\right)\oplus\left(0,\frac{1}{2}\right)$  representation was first proposed by Ahluwalia~\cite{Ahluwalia:2020jkw} and latter studied in~\cite{Ahluwalia:2022zrm}.  Let $\psi(x)$ and $\gdualn{\psi}(x)$ be the massive spinor field and its adjoint. Their expansions are given by
\begin{align}
\psi(x)&=\int\frac{d^{3}p}{(2\pi)^{3}}\frac{1}{\sqrt{2E_{\mathbf{p}}}}\sum_{\sigma=\pm\frac{1}{2}}
\left[e^{-ip\cdot x}u(\p,\sigma)b(\p,\sigma)+e^{ip\cdot x}v(\p,\sigma)b^{c\dag}(\p,\sigma)\right] \, ,\label{eq:a}
\\
\gdualn{\psi}(x)&=\int\frac{d^{3}p}{(2\pi)^{3}}\frac{1}{\sqrt{2E_{\mathbf{p}}}}\sum_{\sigma=\pm\frac{1}{2}}
\left[e^{ip\cdot x}\bar{u}(\p,\sigma)b^{\dag}(\p,\sigma)-e^{-ip\cdot x}\bar{v}(\p,\sigma)b^{c}(\p,\sigma)\right] \,,\label{eq:da}
\end{align}
where $u,v$ are the Dirac spinors and $\bar{u}=u^{\dag}\gamma^{0}$, $\bar{v}=v^{\dag}\gamma^{0}$. At rest $\p=0$, they are given by ref.~\cite{Weinberg:1995mt}
\begin{alignat}{2}
u(\0,\textstyle{\frac{1}{2}})&=\sqrt{m}\left[\begin{matrix}
1 \\
0 \\
1 \\
0
\end{matrix}\right] \, , &&\quad 
u(\0,-\textstyle{\frac{1}{2}})=\sqrt{m}\left[\begin{matrix}
0 \\
1 \\
0 \\
1
\end{matrix}\right] \, ,\\
v(\0,\textstyle{\frac{1}{2}})&=\sqrt{m}\left[\begin{matrix}
0 \\
1 \\
0 \\
-1
\end{matrix}\right] \, , &&\quad
v(\0,-\textstyle{\frac{1}{2}})=\sqrt{m}\left[\begin{matrix}
-1 \\
0 \\
1 \\
0
\end{matrix}\right] \, .
\end{alignat}
At momentum $\p$, the spinors satisfy the Dirac equation
\begin{align}
    (\gamma^{\mu}p_{\mu}-m)u(\p,\sigma)=(\gamma^{\mu}p_{\mu}+m)v(\p,\sigma)=0\,.
\end{align}
Therefore, $\psi(x)$ and $\gdualn{\psi}(x)$ satisfy the Dirac equation
\begin{equation}
    i\gamma^{\mu}\partial_{\mu}\psi(x)-m\psi(x)=-i\partial_{\mu}\gdualn{\psi}(x)\gamma^{\mu}-m\gdualn{\psi}(x)=0\,,
\end{equation}
and have the Lagrangian density
\begin{equation}
    \mathcal{L}_{\psi}=\gdualn{\psi}(i\gamma^{\mu}\partial_{\mu}-m)\psi\,.\label{eq:L_spinor}
\end{equation}
The spinor fields have spin $j=\frac{1}{2}$, so they are bosonic. The equal-time commutator between $\psi(t,\x)$ and $\gdualn{\psi}(t,\y)$ is given by
\begin{equation}
    \left[\psi(t,\x),\gdualn{\psi}(t,\y)\right]_{-}=\gamma^{0}\delta^{(3)}(\x-\y)\,.
\end{equation}
The conjugate momentum for $\psi(x)$ is $\pi(x)=i\gdualn{\psi}(x)\gamma^{0}$. They satisfy the canonical commutation relations
\begin{align}
    \left[\psi(t,\x),\psi(t,\y)\right]_{-}&=\left[\pi(t,\x),\pi(t,\y)\right]_{-}=0\,,\\
    \left[\psi(t,\x),\pi(t,\y)\right]_{-}&=i\delta^{(3)}(\x-\y)\,.
\end{align}

The $C$, $P$ and $T$ transformations for $b(\p,\sigma)$ and $b^{c\dag}(\p,\sigma)$ are the same as the creation and annihilation operators for the Dirac fermions
\begin{align}
    Cb(\p,\sigma)C^{-1}&=\eta^{*}_{C}b^{c}(\p,\sigma)\,,\\
    Pb(\p,\sigma)P^{-1}&=\eta^{*}_{P}b(-\p,\sigma)\,,\\
    Tb(\p,\sigma)T^{-1}&=\eta^{*}_{T}(-1)^{1/2-\sigma}b(-\p,-\sigma)\,,
\end{align}
and
\begin{align}
    Cb^{c\dag}(\p,\sigma)C^{-1}&=\eta^{c}_{C}b^{\dag}(\p,\sigma)\,,\\
    Pb^{c\dag}(\p,\sigma)P^{-1}&=\eta^{c}_{P}b^{c\dag}(-\p,\sigma)\,,\\
    Tb^{c\dag}(\p,\sigma)T^{-1}&=\eta^{c}_{T}(-1)^{1/2-\sigma}b^{c\dag}(-\p,-\sigma)\,.
\end{align}
The spin-half bosons have the same parity and time-reversal transformations as the Dirac fermions. By choosing the intrinsic parity and time-reversal phases to be odd and even
\begin{align}
    \eta^{*}_{P}&=-\eta^{c}_{P}\,,\\
    \eta^{*}_{T}&=\eta^{c}_{T}\,,
\end{align}
we obtain
\begin{alignat}{2}
    P\psi(x)P^{-1}&=\eta^{*}_{P}\gamma^{0}\psi(\mathscr{P}x)\,,\\
    T\psi(x)T^{-1}&=\eta^{*}_{T}\gamma^{1}\gamma^{3}\psi(\mathscr{T}x)\,.
\end{alignat}
The charge-conjugation symmetry for the spin-half bosons has to be analyzed more carefully because here, $\gdualn{\psi}(x)\neq\bar{\psi}(x)$. Acting $C$ on $\psi(x)$ yields
\begin{align}
    C\psi(x)C^{-1}&=\int\frac{d^{3}p}{(2\pi)^{3}}\frac{1}{\sqrt{2E_{\mathbf{p}}}}\sum_{\sigma=\pm\frac{1}{2}}\left[e^{-ip\cdot x}u(\p,\sigma)\eta^{*}_{C}b^{c}(\p,\sigma)+e^{ip\cdot x}v(\p,\sigma)\eta^{c}_{C}b^{\dag}(\p,\sigma)\right] \nonumber\\
    &=\int\frac{d^{3}p}{(2\pi)^{3}}\frac{1}{\sqrt{2E_{\mathbf{p}}}}\sum_{\sigma=\pm\frac{1}{2}}i\gamma^{2}\left[e^{-ip\cdot x}\eta^{*}_{C}v^{*}(\p,\sigma)b^{c}(\p,\sigma)+e^{ip\cdot x}\eta^{c}_{C}u^{*}(\p,\sigma)b^{\dag}(\p,\sigma)\right]\,,
\end{align}
where we have used the identities
\begin{align}
    u(\p,\sigma)&=i\gamma^{2}v^{*}(\p,\sigma)\,,\\
    v(\p,\sigma)&=i\gamma^{2}u^{*}(\p,\sigma)\,.
\end{align}
For charge-conjugation to be conserved, the Lagrangian density must be invariant under the charge-conjugation transformation. For this reason, we cannot choose the intrinsic phase to be even.\footnote{If $\eta^{*}_{C}=\eta^{c}_{C}$, then $C[\gdualn{\lambda}(x)\lambda(x)]C^{-1}=[\gdualn{\lambda}(x)\lambda(x)]^{\dag}$. This is not the correct symmetry for the spinor field because the Lagrangian density is not invariant under charge-conjugation $C\mathcal{L}_{\lambda}C^{-1}=\mathcal{L}^{\dag}_{\lambda}$.} 
Instead, the intrinsic phase must be odd
\begin{equation}
    \eta^{*}_{C}=-\eta^{c}_{C}\,.
\end{equation}
We then obtain
\begin{align}
    C\psi(x)C^{-1}&=i\eta^{*}_{C}\gamma^{2}\gamma^{0}\gdualn{\psi}^{T}(x)\,,\\
    C\gdualn{\psi}(x)C^{-1}&=-i\eta_{C}\psi^{T}(x)\gamma^{0}\gamma^{2}\,,
\end{align}
and $C\gdualn{\psi}(x)\psi(x) C^{-1}=\gdualn{\psi}(x)\psi(x)$. This is the correct charge-conjugation symmetry for the spin-half boson because $C\mathcal{L}_{\psi}(x)C^{-1}=\mathcal{L}_{\psi}(x)$.


\subsection{Vector field}

We now construct the fermionic vector field in the $\left(\frac{1}{2},\frac{1}{2}\right)$ representation. This representation is reducible as it can be decomposed into a direct sum of spin-zero and spin-one irreducible representations. The former is a spin-zero vector field. Here, we focus on the spin-one vector field.
Let $v^{\mu}(x)$ be a complex vector field and $\gdualn{v}^{\mu}(x)$ be its adjoint with the expansions
\begin{align}
    v^{\mu}(x)&=\int\frac{d^{3}p}{(2\pi)^{3}}\frac{1}{\sqrt{2E_{\mathbf{p}}}}\sum_{\sigma=\pm1,0}\left[e^{-ip\cdot x}e^{\mu}(\p,\sigma)b(\p,\sigma)+e^{ip\cdot x}e^{\mu*}(\p,\sigma)b^{c\dag}(\p,\sigma)\right]\,,\\
    \gdualn{v}^{\mu}(x)&=\int\frac{d^{3}p}{(2\pi)^{3}}\frac{1}{\sqrt{2E_{\mathbf{p}}}}\sum_{\sigma=\pm1,0}\left[e^{ip\cdot x}e^{\mu*}(\p,\sigma)b^{\dag}(\p,\sigma)-e^{-ip\cdot x}e^{\mu}(\p,\sigma)b^{c}(\p,\sigma)\right]\,.
\end{align}
In the rest frame $\p=\0$, the coefficients are~\cite{Weinberg:1995mt}
\begin{alignat}{2}
e^{0}(\0,0)&=0\,,&\quad e^{0}(\0,\pm1)&=0\,,\\
e^{1}(\0,0)&=0\,,&\quad e^{1}(\0,\pm1)&=\mp\frac{1}{\sqrt{2}}\,,\\
e^{2}(\0,0)&=0\,,&\quad e^{2}(\0,\pm1)&=-\frac{i}{\sqrt{2}}\,,\\
e^{3}(\0,0)&=1\,,&\quad e^{3}(\0,\pm1)&=0\,.
\end{alignat}
The coefficients at momentum $\p$ are given by 
\begin{equation}
e^{\mu}(\p,\sigma)={L^{\mu}}_{\nu}(\p)e^{\nu}(\0,\sigma)\,,    
\end{equation}
where $L(\p)$ is the Lorentz boost. The momentum $p^{\mu}$ is on the mass-shell $p^{2}=m^{2}$, so the coefficients satisfy the Klein-Gordon equation. The coefficients also satisfy $p_{\mu}e^{\mu}(\p,\sigma)=0$. Therefore, the vector fields satisfy the Klein-Gordon equation
\begin{equation}
    (\partial^{\mu}\partial_{\mu}+m^{2})v^{\mu}(x)=(\partial^{\mu}\partial_{\mu}+m^{2})\gdualn{v}^{\mu}(x)=0\,.
\end{equation}
and the identity
\begin{equation}
    \partial_{\mu}v^{\mu}(x)=\partial_{\mu}\gdualn{v}^{\mu}(x)=0\,.
\end{equation}

The Lagrangian density for the vector field is given by
\begin{equation}
    \mathcal{L}_{v}=-\frac{1}{2}\gdualn{f}^{\mu\nu}f_{\mu\nu}+m^{2}\gdualn{v}^{\mu}v_{\mu} \, ,
\end{equation}
where $f^{\mu\nu}(x)$ and $\gdualn{f}^{\mu\nu}(x)$ are the respective field strength tensors for $v^{\mu}(x)$ and $\gdualn{v}^{\mu}(x)$
\begin{align}
    \gdualn{f}^{\mu\nu}&=\partial^{\mu}\gdualn{v}^{\nu}-\partial^{\nu}\gdualn{v}^{\mu}\,,\\
    f_{\mu\nu}&=\partial_{\mu}v_{\nu}-\partial_{\nu}v_{\mu}\,.
\end{align}
The equations of motion and conjugate momenta are given by
\begin{align}
    \partial_{\mu}f^{\mu\nu}+m^{2}v^{\nu}=
    \partial_{\mu}\gdualn{f}^{\mu\nu}+m^{2}\gdualn{v}^{\nu}=0 \, ,\label{eq:eom_vector}
\end{align}
and
\begin{align}
    \pi_{\mu}=\frac{\partial\mathcal{L}_{v}}{\partial\dot{v}^{\mu}}=\gdualn{f}_{\mu0}\,,\quad
    \gdualn{\pi}_{\mu}=\frac{\partial\mathcal{L}_{v}}{\partial\dot{\gdualn{v}}^{\mu}}=-f_{\mu0}\,.\label{eq:pi}
\end{align}

The vector fields are of spin $j=1$, so they are fermionic. To compute the canonical anti-commutators, we first have to identify the independent components of the vector fields and their conjugate momenta. From eq.~\eqref{eq:pi}, we find $\pi_{0}(x)=\gdualn{\pi}_{0}(x)=0$. The non-vanishing, independent components of the conjugate momenta are
\begin{align}
    \pi_{i} &=\gdualn{f}_{i0} = \partial_{i}\gdualn{v}_{0}-\dot{\gdualn{v}}_{i}\,,\\
    \gdualn{\pi}_{i} &=-f_{i0} =-\partial_{i}v_{0}+\dot{v}_{i}\,.\label{eq:d_pi}
\end{align}
From the equations of motion~\eqref{eq:eom_vector} for the $\nu=0$ component, we find that $v^{0}(x)$ and $\gdualn{v}^{0}(x)$ can be expressed in terms of the divergence of the conjugate momenta
\begin{align}
    v^{0}(x)=\frac{1}{m^{2}}g^{ij}\partial_{i}\gdualn{\pi}_{j}(x)\,,\quad
    \gdualn{v}^{0}(x)=-\frac{1}{m^{2}}g^{ij}\partial_{i}\pi_{j}(x)\,.
\end{align}
Therefore, the independent components of the vector fields and conjugate momenta are $v^{i},\gdualn{v}^{i}$ and $\pi_{j},\gdualn{\pi}_{j}$. From eq.~\eqref{eq:d_pi}, $\pi_{j}$ and $\gdualn{\pi}_{j}$ are given by
\begin{align}
    \pi_{j}(x)&=i\int\frac{d^{3}p}{(2\pi)^{3}}\frac{1}{\sqrt{2E_{\mathbf{p}}}}\sum_{\sigma=\pm1,0}\left[e^{ip\cdot x}f^{*}_{j0}(\p,\sigma)b^{\dag}(\p,\sigma)+e^{-ip\cdot x}f_{j0}(\p,\sigma)b^{c}(\p,\sigma)\right] \, ,\\
    \gdualn{\pi}_{j}(x)&=i\int\frac{d^{3}p}{(2\pi)^{3}}\frac{1}{\sqrt{2E_{\mathbf{p}}}}\sum_{\sigma=\pm1,0}\left[e^{-ip\cdot x}f_{j0}(\p,\sigma)b(\p,\sigma)-e^{ip\cdot x}f^{*}_{j0}(\p,\sigma)b^{c\dag}(\p,\sigma)\right] \, ,
\end{align}
where
\begin{equation}
    f_{j0}(\p,\sigma)=p_{j}e_{0}(\p,\sigma)-p_{0}e_{j}(\p,\sigma)\,,\quad p_{0}=E_{\mathbf{p}}\,.
\end{equation}
The canonical equal-time anti-commutators between the vector fields and their conjugate momenta are given by
\begin{align}
    \left[v^{i}(t,\x),v^{j}(t,\y)\right]_{+}&=0\,,\\
    \left[\pi_{i}(t,\x),\pi_{j}(t,\y)\right]_{+}&=0\,,\\
    \left[v^{i}(t,\x),\pi_{j}(t,\y)\right]_{+}&=i\delta^{i}_{j}\delta^{(3)}(\x-\y)\,,
\end{align}
and
\begin{align}
    \left[\gdualn{v}^{i}(t,\x),\gdualn{v}^{j}(t,\y)\right]_{+}&=0\,,\\
    \left[\gdualn{\pi}_{i}(t,\x),\gdualn{\pi}_{j}(t,\y)\right]_{+}&=0\,,\\
    \left[\gdualn{v}^{i}(t,\x),\gdualn{\pi}_{j}(t,\y)\right]_{+}&=i\delta^{i}_{j}\delta^{(3)}(\x-\y)\,.
\end{align}
The remaining anti-commutators identically vanish
\begin{align}
    \left[v^{i}(t,\x),\gdualn{v}^{j}(t,\y)\right]_{+}&=\left[\pi_{i}(t,\x),\gdualn{\pi}_{j}(t,\y)\right]_{+}=0\,,\nonumber\\
    \left[v^{i}(t,\x),\gdualn{\pi}_{j}(t,\y)\right]_{+}&=\left[\gdualn{v}^{i}(t,\x),\pi_{j}(t,\y)\right]_{+}=0\,.
\end{align}

The discrete symmetry transformations for $b(\p,\sigma)$ and $b^{c\dag}(\p,\sigma)$ are
\begin{align}
    Cb(\p,\sigma)C^{-1}&=\eta^{*}_{C}b^{c}(\p,\sigma)\,,\\
    Pb(\p,\sigma)P^{-1}&=\eta^{*}_{P}b(-\p,\sigma)\,,\\
    Tb(\p,\sigma)T^{-1}&=\eta^{*}_{T}(-1)^{1-\sigma}b(-\p,-\sigma)\,,
\end{align}
and
\begin{align}
    Cb^{c\dag}(\p,\sigma)C^{-1}&=\eta^{c}_{C}b^{\dag}(\p,\sigma)\,,\\
    Pb^{c\dag}(\p,\sigma)P^{-1}&=\eta^{c}_{P}b^{c\dag}(-\p,\sigma)\,,\\
    Tb^{c\dag}(\p,\sigma)T^{-1}&=\eta^{c}_{T}(-1)^{1-\sigma}b^{c\dag}(-\p,-\sigma)\,.
\end{align}
For  parity and time-reversal transformations, we choose the intrinsic phases to be even
\begin{align}
   \eta^{*}_{P}&=\eta^{c}_{P}\,,\\
    \eta^{*}_{T}&=\eta^{c}_{T}\,,
\end{align}
to obtain
\begin{alignat}{2}
    Pv^{\mu}(x)P^{-1}&=-\eta^{*}_{P}{\mathscr{P}^{\mu}}_{\nu}v^{\nu}(\mathscr{P}x)\,,\\
    Tv^{\mu}(x)T^{-1}&=-\eta^{*}_{T}{\mathscr{T}^{\mu}}_{\nu}v^{\nu}(\mathscr{T}x)\,.
\end{alignat}
Here, charge-conjugation maps $v^{\mu}(x)$ to $\gdualn{v}^{\mu}(x)$ instead of $v^{\mu\dag}(x)$. Therefore, by choosing the intrinsic charge-conjugation phases to be odd
\begin{equation}
      \eta^{*}_{C}=-\eta^{c}_{C}\,,\\
\end{equation}
we obtain
\begin{equation}
        Cv^{\mu}(x)C^{-1}=\eta^{*}_{C}\gdualn{v}^{\mu}(x)\,.
\end{equation}

\section{Poincar\'{e} symmetry \& Belinfante-Rosenfeld tensor}\label{sec:Lagrangian}



The pseudo-Hermitian fields constructed in the previous section preserve Poincar\'{e} symmetry. Consequently, their Lagrangian densities are invariant under a set of global Lorentz transformations from which we can, by Noether’s theorem, derive the generators of the Poincar\'{e} group. Although these generators resemble their Hermitian counterparts, they exhibit a crucial difference: \textit{the free generators are Hermitian and satisfy the Poincar\'{e} algebra, but the Belinfante-Rosenfeld tensor is pseudo-Hermitian.} We prove these results below.

Let $\chi(x)$ and $\gdualn{\chi}(x)$ be the free canonical field and its pseudo-Hermitian conjugate of~eqs.~\eqref{eq:chi}-\eqref{eq:chi_adj}. The Lagrangian density depends on $\chi(x),\gdualn{\chi}(x)$ and their derivatives. The theory respects Poincar\'{e} symmetry, so the Lagrangian density is invariant under the global transformations
\begin{align}
    \delta \chi(x)&=+\frac{i}{2}\omega^{\mu\nu}\mathcal{J}_{\mu\nu}\chi(x)\,,\\
    \delta\gdualn{\chi}(x)&=-\frac{i}{2}\gdualn{\chi}\omega^{\mu\nu}\mathcal{J}_{\mu\nu}(x)\,,
\end{align}
where $\mathcal{J}$ is the finite-dimensional generator of the Lorentz group and $\omega$ is the Lorentz transformation parameter. Their derivatives transform as
\begin{align}
    \delta\left[\partial_{\kappa}\chi(x)\right]&=+\frac{i}{2}\omega^{\mu\nu}\mathcal{J}_{\mu\nu}\partial_{\kappa}\chi(x)+{\omega_{\kappa}}^{\lambda}\partial_{\lambda}\chi(x)\,,\\
    \delta\big[\partial_{\kappa}\gdualn{\chi}(x)\big]&=-\frac{i}{2}\partial_{\kappa}\gdualn{\chi}(x)\omega^{\mu\nu}\mathcal{J}_{\mu\nu}+{\omega_{\kappa}}^{\lambda}\partial_{\lambda}\gdualn{\chi}(x)\,.
\end{align}
Using the canonical methods~\cite[sec.~7]{Weinberg:1995mt}, we obtain the Belinfante-Rosenfeld tensor
\begin{align}
    \Theta^{\mu\nu}=T^{\mu\nu}&-\frac{i}{2}\partial_{\kappa}\left[\frac{\partial\mathcal{L}}{\partial(\partial_{\kappa}\chi)}\mathcal{J}^{\mu\nu}\chi-\frac{\partial\mathcal{L}}{\partial(\partial_{\mu}\chi)}\mathcal{J}^{\kappa\nu}\chi-\frac{\partial\mathcal{L}}{\partial(\partial_{\nu}\chi)}\mathcal{J}^{\kappa\nu}\chi\right]\nonumber\\
    &+\frac{i}{2}\partial_{\kappa}\left[\gdualn{\chi}\mathcal{J}^{\mu\nu}\frac{\partial\mathcal{L}}{\partial(\partial_{\kappa}\gdualn{\chi})}-\gdualn{\chi}\mathcal{J}^{\kappa\nu}\frac{\partial\mathcal{L}}{\partial(\partial_{\mu}\gdualn{\chi})}-\gdualn{\chi}\mathcal{J}^{\kappa\nu}\frac{\partial\mathcal{L}}{\partial(\partial_{\nu}\gdualn{\chi})}\right]\,,
\end{align}
where 
\begin{equation}
    T^{\mu\nu}=-\eta^{\mu\nu}\mathcal{L}+\frac{\partial\mathcal{L}}{\partial(\partial_{\mu}\chi)}\partial^{\nu}\chi+\partial^{\nu}\gdualn{\chi}\frac{\partial\mathcal{L}}{\partial(\partial_{\mu}\gdualn{\chi})}\,.   
\end{equation}
The Hamiltonian $H_{0}$ and three-momentum $\P_{0}$ are given by
\begin{align}
    H_{0}&=\int d^{3}x\,\Theta^{00}\,,\\
    P^{i}_{0}&=\int d^{3}x\,\Theta^{0i}\,.
\end{align}
From the Belinfante-Rosenfeld tensor, we can construct a third-ranked tensor
\begin{equation}
    M^{\lambda\mu\nu}\equiv x^{\mu}\Theta^{\lambda\nu}-x^{\nu}\Theta^{\lambda\mu}\,,
\end{equation}
to obtain the Lorentz generator
\begin{equation}
    J^{\mu\nu}_{0}=\int d^{3}x\,M^{0\mu\nu}\,.
\end{equation}

The Belinfante-Rosenfeld tensor and the generators of the Poincar\'{e} group are manifestly pseudo-Hermitian with respect to $\eta$. However, it is unclear whether they are Hermitian. Here, we show that the generators of the Poincar\'{e} group are Hermitian, but the Belinfante-Rosenfeld tensor is non-Hermitian. To prove the Hermiticity of the generators, let us consider the relations between the pseudo-Hermitian and Hermitian QFTs. In a given irreducible representation of the Poincar\'{e} group, furnished by both $\chi,\gdualn{\chi}$ and $\lambda,\bar{\lambda}$, they are related by
\begin{align}
    \chi(x)&=\lambda(x)\vert_{a\rightarrow b,\,a^{c\dag}\rightarrow b^{c\dag}}\,,\label{eq:chi_lambda}\\
    \gdualn{\chi}(x)&=\bar{\lambda}(x)\vert_{a^{\dag}\rightarrow b^{\dag},\,a^{c}\rightarrow-b^{c}}\,.\label{eq:dchi_dlambda}
\end{align}
The Hermitian QFTs are well-defined in the sense that their $H_{0}$ and $\mathbf{P}_{0}$ are Hermitian 
\begin{align}
    H_{0}(\lambda)&=\int\frac{d^{3}p}{(2\pi)^{3}}\,E_{\mathbf{p}}\sum_{\sigma}\left[a^{\dag}(\p,\sigma)a(\p,\sigma)\pm a^{c}(\p,\sigma)a^{c\dag}(\p,\sigma)\right]\,,\label{eq:H0}\\
    \mathbf{P}_{0}(\lambda)&=\int\frac{d^{3}p}{(2\pi)^{3}}\,\p\sum_{\sigma}\left[a^{\dag}(\p,\sigma)a(\p,\sigma)\pm a^{c}(\p,\sigma)a^{c\dag}(\p,\sigma)\right]\,,\label{eq:P0}
\end{align}
where the top and bottom signs in eqs.~\eqref{eq:H0}-\eqref{eq:P0} are for the bosonic ($j=0,1,\cdots$) and fermionic fields ($j=\frac{1}{2},\frac{3}{2},\cdots$). From eqs.~\eqref{eq:chi_lambda}-\eqref{eq:dchi_dlambda}, it follows that $H_{0}$ and $\mathbf{P}_{0}$ for pseudo-Hermitian QFTs are also Hermitian~
\begin{align}
    H_{0}(\chi)&=\int\frac{d^{3}p}{(2\pi)^{3}}\,E_{\mathbf{p}}\sum_{\sigma}\left[b^{\dag}(\p,\sigma)b(\p,\sigma)\mp b^{c}(\p,\sigma)b^{c\dag}(\p,\sigma)\right]\,,\label{eq:H0_pH}\\
    \mathbf{P}_{0}(\chi)&=\int\frac{d^{3}p}{(2\pi)^{3}}\,\p\sum_{\sigma}\left[b^{\dag}(\p,\sigma)b(\p,\sigma)\mp b^{c}(\p,\sigma)b^{c\dag}(\p,\sigma)\right]\,.\label{eq:P0_pH}
\end{align}
But this time, the top and bottom signs apply to fermionic ($j=0,1,\cdots$) and bosonic fields ($j=\frac{1}{2},\frac{3}{2},\cdots$), respectively. For the Lorentz generator, we do not need the expression for $J^{\rho\sigma}(\chi)$ in terms of the creation and annihilation operators to prove its Hermiticity. Knowing that $H_{0}(\chi)$ and $\mathbf{P}_{0}(\chi)$ are Hermitian and that the Poincar\'{e} generators for $\chi,\gdualn{\chi}$ derived from the canonical formalism satisfy the Poincar\'{e} algebra suffice~\cite[Sec.~7]{Weinberg:1995mt}. To show that the Lorentz generator $J^{\rho\sigma}_{0}(\chi)$ is Hermitian, we consider its commutator with the momentum operator $P^{\mu}_{0}(\chi)$ where $P^{0}_{0}(\chi)=H_{0}(\chi)$
\begin{equation}
    i[P^{\mu}_{0}(\chi),J^{\rho\sigma}_{0}(\chi)]=-g^{\mu\rho}P^{\sigma}_{0}(\chi)+g^{\mu\sigma}P^{\rho}_{0}(\chi)\,.\label{eq:pj}
\end{equation}
Taking the Hermitian-conjugate of eq.~\eqref{eq:pj}, we obtain
\begin{equation}
    i[P^{\mu}_{0}(\chi),J^{\rho\sigma\dag}_{0}(\chi)]=-g^{\mu\rho}P^{\sigma}_{0}(\chi)+g^{\mu\sigma}P^{\rho}_{0}(\chi)\,.
\end{equation}
Since $P^{\mu}_{0}(\chi)=P^{\mu\dag}_{0}(\chi)$, the Lorentz generator must also be Hermitian, $J^{\rho\sigma}_{0}(\chi)=J^{\rho\sigma\dag}_{0}(\chi)$.

The fact that the Poincar\'{e} generators are Hermitian is a testament that the free pseudo-Hermitian QFT is unitary. However, this does not change the fact that the Belinfante-Rosenfeld tensor is non-Hermitian. To see this, let us consider the symplectic fermion, whose Belinfante-Rosenfeld tensor is\footnote{If we take the symplectic fermion as an elementary particle, then it must couple to gravity. To the leading order in Minkowski spacetime, the gravitational interaction takes the form $\int d^{3}x\,\Theta^{\mu\nu}_{\phi}h_{\mu\nu}$ with $h_{\mu\nu}$ being the gravitational field, which is non-Hermitian.}
\begin{equation}
\Theta^{\mu\nu}_{\phi}=-g^{\mu\nu}\left(\partial^{\rho}\gdualn{\phi}\partial_{\rho}\phi-m^{2}\gdualn{\phi}\phi\right)+\partial^{\mu}\gdualn{\phi}\partial^{\nu}\phi+\partial^{\nu}\gdualn{\phi}\partial^{\mu}\phi 
\, .
\label{eq:Theta_phi}
\end{equation}
Each term in eq.~\eqref{eq:Theta_phi} is non-Hermitian. Nevertheless, direct computation shows that $\int d^{3}x\,\Theta^{0\mu}_{\phi}$ is Hermitian. Similarly, $\int d^{3}x\,\Theta^{ij}_{\phi}$ is non-Hermitian.

\section{Conclusion} \label{sec:conclusion}

The pseudo-Hermitian QFT provides a framework for generalizing conventional unitary QFT. Here, we extend the formalism developed by LeClair et al.~\cite{LeClair_2007,Robinson:2009xm} and Ahluwalia et al.~\cite{Ahluwalia:2020jkw,Ahluwalia:2022zrm}. to construct a class of free pseudo-Hermitian QFTs that obey the NSST. Within this framework, integer-spin fields are fermionic and half-integer-spin fields are bosonic. The generators of the Poincar\'{e} group are Hermitian, ensuring that the free field theory is unitary.

Our construct is not motivated by particle phenomenology, so it does not directly address the well-known problems related to the Standard Model. Instead, we explored the foundations of QFT and unearthed a hidden structure in the field adjoint. This discovery naturally yields a new class of pseudo-Hermitian QFTs obeying the NSST. The question is: How do we make sense of the pseudo-Hermitian fields? We have shown that for every charged bosonic/fermionic Hermitian QFT, we can construct a fermionic/bosonic pseudo-Hermitian QFT of equal mass and spin. This is similar to supersymmetry except that here, the bosons and fermions have equal mass and spin. If the pseudo-Hermitian fields describe elementary particles, they must be new particles beyond the Standard Model. To predict the signatures for pseudo-Hermitian fields, it is necessary to study the interacting theories.

In the absence of interactions, the free pseudo-Hermitian QFTs are unitary and physically well-defined. However, in order for these fields to describe elementary particles, they must have consistent interacting theories. Formulating a consistent interacting pseudo-Hermitian QFT is a major challenge that we have to overcome.
In the presence of interactions, the full Hamiltonian becomes non-Hermitian while being pseudo-Hermitian with respect to $\eta$ where $\eta^{-1}H^{\dag}\eta=H$. For states whose time evolution are given by $|\alpha,t\rangle=e^{-iHt}|\alpha\rangle$ and $\langle\beta,t|=\langle\beta|e^{iH^{\dag}t}$, the time translation symmetry can be preserved by defining the $\eta$-inner product \'{a} la Mostafazadeh, namely $\langle\beta|\alpha\rangle_{\eta}\equiv\langle\beta|\eta|\alpha\rangle$. Using the fact that $H$ is pseudo-Hermitian, we find $\langle\beta,t|\alpha,t\rangle_{\eta}=\langle\beta|\alpha\rangle_{\eta}$. Within this framework, when the full Hamiltonian $H$ and the free Hamiltonian $H_{0}$ have the same spectrum, the resulting $S$-matrix satisfies the generalized unitary relation $\eta^{-1}S^\dagger \eta=S^{-1}$~\cite{Lee:2023aip}. However, the inner product defined via $\eta$ does not suffice because it is not positive-definite (the eigenvalue of $\eta$ is $\pm1$). An interacting theory in agreement with the canonical probabilistic interpretation of quantum mechanics requires a positive-definite $\eta_{+}$-inner product. Constructing a positive-definite $\eta_{+}$ for pseudo-Hermitian scattering theory is an important topic for future investigation.



\appendix

\section*{Appendix}

\section{Notations \& conventions}
\label{app:notations_and_conventions}

Throughout the paper, we use the conventions of ref.~\cite{Peskin:1995ev}.
The 4D Minkowski metric is,
\begin{equation}
g_{\mu \nu} = g^{\mu \nu} = \text{diag}(+1, -1, -1, -1) \, ,
\label{metric_1}
\end{equation}
where $\mu, \nu=0,1, 2, 3$.

The $4\times4$ Dirac matrices are taken in the Weyl representation,
\begin{equation}
\gamma^\mu =
\left[\begin{matrix}
0 & \sigma^\mu \\
\bar{\sigma}^\mu & 0
\end{matrix}\right]
\, ,
\quad \text{with} \quad
\biggl\{
\begin{array}{r c l}
\sigma^\mu &=& \left( \mathds{1}_{2 \times 2}, \sigma^i \right) \, , \\
\bar{\sigma}^\mu &=& \left( \mathds{1}_{2 \times 2}, -\sigma^i \right) \, ,
\end{array}
\label{gamma_1}
\end{equation}
where $\mu=0,1,2,3$ and $\sigma^i$ ($i = 1, 2, 3$) are the three Pauli matrices:
\begin{equation}
\sigma^1 =
\left[\begin{matrix}
0 \quad  & \quad 1  \\
1 \quad  & \quad 0  
\end{matrix}\right] \, ,
\quad
\sigma^2 =
\left[\begin{matrix}
0 \quad  & -i \\
i \quad  & 0
\end{matrix}\right] \, ,
\quad
\sigma^3 =
\left[\begin{matrix}
1 \quad  &  0 \\
0 \quad  &  -1
\end{matrix}\right] \, ,
\end{equation}
and
\begin{align}
\left\{\sigma^{i}, \sigma^{j} \right\} = 2\delta^{ij} \, , \quad i,j = 1 \cdots 3 \, ,
\end{align}
where the Kronecker delta function is
\begin{equation}
\delta^{ij} = 
\biggl\{
\begin{array}{r c l}
&1&, \  i=j\\
&0&, \  i \neq j
\end{array} \, .
\end{equation}
The Levi-Civita tensor $\epsilon_{ijk}$, where $i,j,k = 1 \cdots 3$ is totally antisymmetric, with $\epsilon_{123}=+1$.

One also has the chiral operator,
\begin{equation}
\gamma^5 = i \gamma^0 \gamma^1 \gamma^2 \gamma^3 =
\left[\begin{matrix}
- \mathds{1}_{2 \times 2} & 0 \\
0 & \mathds{1}_{2 \times 2}
\end{matrix}\right] \, ,
\label{gamma_2}
\end{equation}
and the chiral projection operators
\begin{equation}
P_{L,R} \equiv \dfrac{\mathds{1} \mp \gamma^{5}}{2} \, .
\end{equation}

\acknowledgments

We would like to thank Ting-Long Feng for collaboration at the early stage of this work. SZ is supported by the Natural Science Foundation of China under Grant No.12347101, No.2024CDJXY022 and No.CSTB2024YCJH-KYXM0070 at Chongqing University.

\bibliography{Bibliography}
\bibliographystyle{JHEP}

\end{document}